\documentclass{aa}

\usepackage{graphicx}
\usepackage{txfonts}
\usepackage{url}
\usepackage{wasysym}
\usepackage{longtable}
\usepackage{array}
\usepackage{tikz}
\usepackage[pdftitle={Exocomet orbital distribution around $\beta$\,Pictoris}, colorlinks = true, breaklinks = true, citecolor = blue, linkcolor = blue, urlcolor = blue, pdfauthor = {Ren\'{e} Heller}]{hyperref}
\usepackage{esvect}  

\defcitealias{LdE2022}{L$+$22}
\newcommand{\LecDE}{\citetalias{LdE2022}}

\begin{document} 

\title{Exocomet orbital distribution around $\beta$\,Pictoris}
\titlerunning{Exocomet orbital distribution around $\beta$\,Pictoris}

\author{
  Ren{\'e} Heller\inst{1}
}

\institute{
Max-Planck-Institut f\"ur Sonnensystemforschung, Justus-von-Liebig-Weg 3, 37077 G\"ottingen, Germany; 
\href{mailto:heller@mps.mpg.de}{heller@mps.mpg.de}
}

   \date{Received 23 May 2022 / Accepted 30 July 2024}

 
\abstract{
The ${\sim}23$\,Myr young star $\beta$\,Pictoris ($\beta$\,Pic) is a laboratory for planet formation studies given its observed {debris} disk, its directly imaged super-Jovian planets $\beta$\,Pic\,b and c, and the evidence of extrasolar comets regularly transiting in front of the star. The most recent evidence of exocometary transits around $\beta$\,Pic came from stellar photometric time series obtained with the TESS space mission. Previous analyses of these transits constrained the orbital distribution of the underlying exocomet population to range between about 0.03 and 1.3\,AU assuming a fixed transit impact parameter. Here we examine the distribution of the observed transit durations (${\Delta}t$) to infer the orbital {surface} density distribution ($\delta$) of the underlying exocomet sample. The effect of the geometric transit probability for circular orbits is properly taken into account but it is assumed that the radius of the transiting comets and their possible clouds of evaporating material are much smaller than the stellar radius. We show that a narrow belt of exocomets around $\beta$\,Pic, in which the transit impact parameters are randomized but the orbital semimajor axes are equal, results in a pile-up of long transit durations. This is contrary to observations, which reveal a pile-up of short transit durations (${\Delta}t \approx 0.1$\,d) and a tail of only a few transits with ${\Delta}t > 0.4$\,d. A flat density distribution of exocomets between about 0.03 and 2.5\,AU results in a better match between the resulting ${\Delta}t$ distribution and the observations but the slope of the predicted ${\Delta}t$ histogram is not sufficiently steep. {An even better} match to the observations can be produced with a $\delta \propto a^{\beta}$ power law. Our modeling reveals a best fit between the observed and predicted ${\Delta}t$ distribution for $\beta = -0.15_{-0.10}^{+0.05}$. {A more reasonable scenario in which the exocometary trajectories are modeled as hyperbolic orbits can also reproduce the observed ${\Delta}t$ distribution to some extent. Future studies might reproduce the observed ${\Delta}t$ distribution with a full exploration of the parameter four-dimensional space of highly eccentric orbits and they might need to relax our assumption that the transiting objects are smaller than the stellar disk.} The number of observed exocometary transits around $\beta$\,Pic is currently too small to validate the previously reported distinction of two distinct exocomet families but this might be possible with future TESS observations of this star. Our results do nevertheless imply that cometary material exists on highly eccentric orbits with a more extended range of semimajor axes than suggested by previous spectroscopic observations.}


   \maketitle
%

\section{Introduction}

The star $\beta$\,Pictoris ($\beta$\,Pic) has become one of the few benchmark systems for studies of planet formation, in which a {debris} disk \citep{1984Sci...226.1421S}, extrasolar comets \cite{1987A&A...185..267F,1990A&A...236..202B,2014Natur.514..462K}, and even two extrasolar giant planets ($\beta$\,Pic\,b \citealt{2009A&A...493L..21L,2010Sci...329...57L} and c \citealt{2019NatAs...3.1135L,2020A&A...642L...2N}) have been found. It is a naked-eye ($m_V = 3.86$) A-type star on the zero-age main sequence. Its parallax measurement of $50.903\,({\pm}\,0.1482)$ milliarcseconds from Gaia ERD3 \citep{2016A&A...595A...1G,2021A&A...649A...1G} suggests a distance of $19.6345_{-0.0569}^{+0.0573}$\,pc. As part of a co-moving group of stars, the age of $\beta$\,Pic has become exceptionally well constrained. Isochronal fitting of a total of 30 A-, F, and G-type stars from the $\beta$\,Pic moving group suggest an age of $23\,({\pm}3)$\,Myr \citep{2014MNRAS.445.2169M}, and more modern dynamical age estimates based on Gaia DR2 yield $18.5_{-2.4}^{+2.0}$\,Myr \citep{2020A&A...642A.179M}.

The detection and characterization of exocomets around $\beta$\,Pic has an exceptionally long history, with an initial discovery preceding the detection of the first exoplanets by almost a decade. First hints to exocomets came from variations in the Ca\,\textsc{ii}-K absorption line profile of $\beta$\,Pic that had been interpreted as evidence of infalling cometary material on the star \citep{1987A&A...185..267F,1990A&A...236..202B}. More recent, high-resolution stellar spectroscopy distinguished two families of comets located at $10\,({\pm}\,3)\,R_{\rm s}$ and $19\,({\pm}\,4)\,R_{\rm s}$ \cite{2014Natur.514..462K}, respectively, where $R_{\rm s}$ is the stellar radius.

In addition to hints from spectroscopy, variations in the apparent brightness of $\beta$\,Pic have been noted in ground-based archival data from La Silla by the Geneva Observatory. A peculiar dimming event around 10 November 1981, which took several days, has been interpreted as the passage of a planet or a group of planets or planetesimals with a total cross section area comparable to that of Jupiter's \citep{1995A&A...299..557L}. {That being said, recent follow-up observations of $\beta$\,Pic during the transit of the outer regions of the Hill sphere of its giant exoplanet $\beta$\,Pic\,b did not reveal any new such dimming events \citep{2021A&A...648A..15K}. Modern} space-based stellar photometry of $\beta$\,Pic from the ongoing TESS mission led to the discovery of three cometary transits \citep{2019A&A...625L..13Z}, a phenomenon that had been anticipated long ago \citep{1999A&A...343..916L}. These three events, observed in TESS Sectors 4--7 from 19 October 2018 to 1 February 2019, were monitored with much better time resolution than the historical data. In fact, TESS' 2-min short cadence of $\beta$\,Pic is key to the characterization of transit duration of exocometary events presented in this report.

More recently, \citet{LdE2022} (\LecDE) analyzed additional TESS observations of $\beta$\,Pic in Sectors 31--34 from 20 November 2020 to 8 February 2021. Some of these transits were also reported by \citet{2022A&A...660A..49P}. In their combined analyses of the entire TESS data of $\beta$\,Pic, \LecDE \, found 30 cometary transits, including the three previously known. Their analysis also revealed a striking similarity between the cumulative size distribution of the comets around $\beta$\,Pic to the cometary size distribution observed in the solar system. Both distributions are consistent with the canonical size distribution first derived by \citet{1969JGR....74.2531D} for a collisionally relaxed population.

{Apart from cometary studies around $\beta$\,Pic, analyses of the space-based photometry of the Kepler mission showed evidence of cometary transits around other early-type stars, such as KIC\,3542116 \citep{2018MNRAS.474.1453R} and KIC\,8462852 \citep{2016MNRAS.457.3988B, 2018MNRAS.473.5286W}. These recent results show that the community is currently transitioning from an era of exocometary transit detections to an era of exocomet characterization.

{Using the distribution of the observed exocomet transit times around $\beta$\,Pic, we attempt such a characterization in this study.} \LecDE \, estimate the {periastron} distances to range between 0.03\,AU and 1.3\,AU around $\beta$\,Pic. {Their calculations} assumed that each comet transits the star across the entire stellar diameter. Here we extend the analysis of the orbital distribution of these exocomets by investigating the histogram of the transit duration variations for random transit impact parameters and taking into account the transit probability as a function of the distance to the star. {First we investigate extended cometary rings around $\beta$\,Pic. But since comets are unlikely to survive evaporation at distances ${\lesssim}1$\,AU for very long, we also investigate comets on highly eccentric ($e{\sim}1$) hyperbolic orbits.}

\section{Methods}

An important assumption throughout this study is that the radii of the transiting comets are smaller than the stellar radius. This assumption is justified by the observed transits durations, as we demonstrate in Sect.~\ref{sec:cometarysize}.

\subsection{Circular orbits}

{For the purpose of illustration, let} us first consider a star that is transited by a sample of exocomets on circular orbits with randomized transit impact parameters ($b$). Figure~\ref{fig:b}(a) displays the resulting transit paths with length

\begin{equation} \label{eq:p}
p = 2R_{\rm s}\sqrt{1-b^2} \ .
\end{equation}

\noindent
If all these objects are arranged in a narrow ring around the star with essentially the same orbital semimajor axis ($a$) but slightly different orbital inclinations with respect to the line of sight ($i$), then they will all have the same orbital velocity but exhibit different transit durations (${\Delta}t$) depending on their respective transit impact parameter. The resulting histogram of such a {hypothetical} narrow-ring exocomet population is shown in Fig.~\ref{fig:b}(b), where we can see a pile-up towards large values of ${\Delta}t$. As a consequence, the shortest transit durations can be expected to be rare if the ${\Delta}t$ distribution were caused by a comet sample with similar orbital distances.

In $\beta$\,Pic, however, this is not the case. Inspection of the ${\Delta}t$ values measured by \LecDE \, (Extended Data Table 2 therein) shows the exact opposite, namely a ${\Delta}t$ pile-up towards low values. Counting the measurements of \LecDE \, in bins of 0.1\,d width results in the histogram shown in Fig.~\ref{fig:LdE}. It is important to note that the lack of objects with long durations is not due to a sensitivity bias that would favor short transit durations. Quite the contrary, transit detectability increases with transit duration. 

\begin{figure}
    \centering
    \includegraphics[width=1\linewidth]{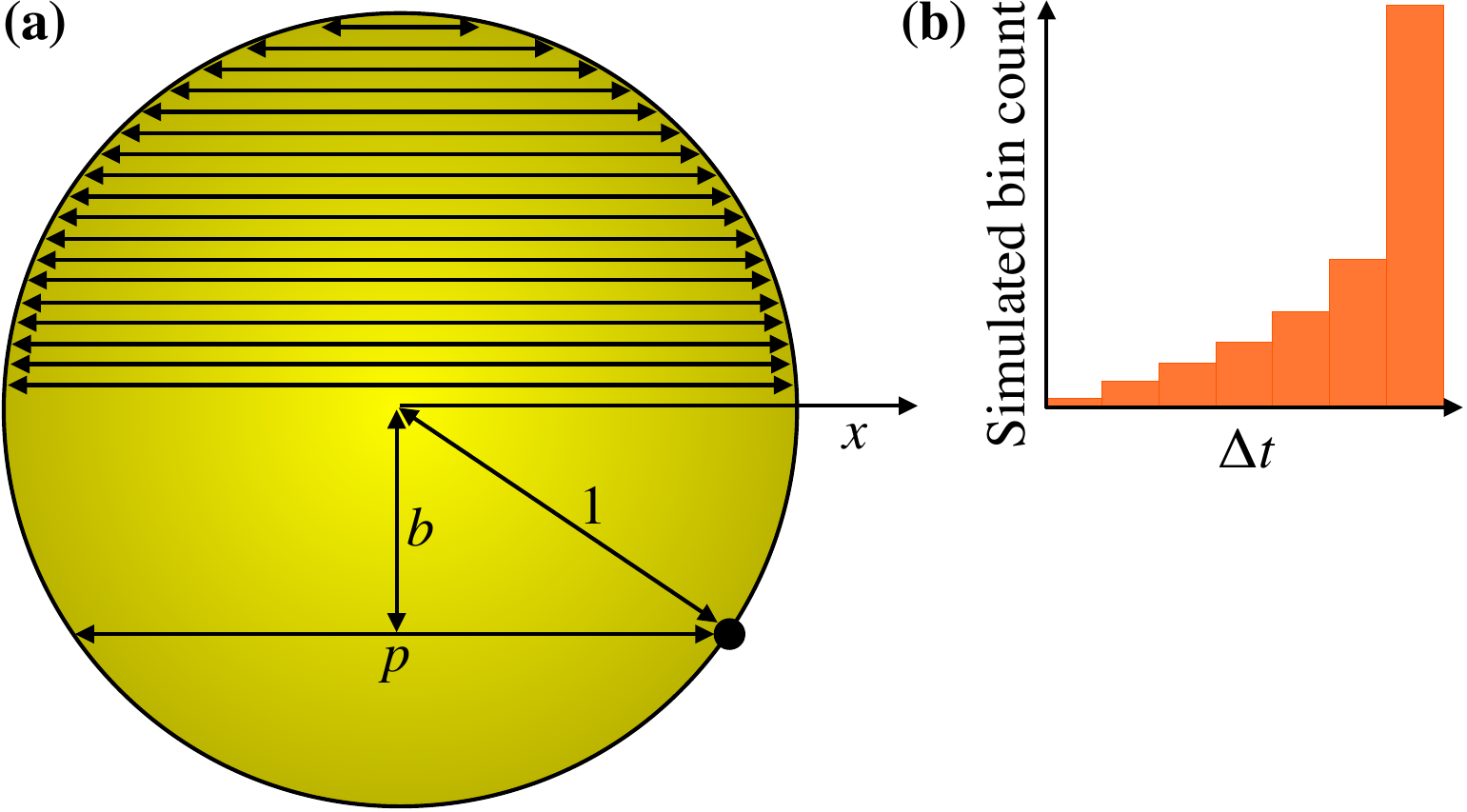}
    \caption{Illustration of the pile-up of long transit paths (and long transit durations, ${\Delta}t$) for randomly distributed transit impact parameters.}
    \label{fig:b}
\end{figure}

Examination of the \LecDE \, data reveals there {are 15 transits within [0.1\,d, 0.2\,d], five of which have the lowest value of ${\Delta}t\,=\,0.10$\,d, and only one transit in each of the ${\Delta}t$ intervals [0.4\,d, 0.5\,d], [0.5\,d, 0.6\,d], and [0.6\,d, 0.7\,d], respectively.} The five short-duration transits are very likely not near-grazing transits because near-grazing transits are expected to be very rare {if $i$ has a flat prior and is randomly distributed} (see Fig.~\ref{fig:b}). Instead, at least some of them are very likely to have transit impact parameters near zero. As a consequence, they should also have the shortest orbital period.

In the limit of circular orbits and under the assumption that the radius of the comets and any possible clouds of evaporating material are much smaller than the stellar radius, the transit duration is given as the transit path divided by the orbital velocity,

\begin{equation}\label{eq:Dt}
{\Delta}t = \frac{2 R_{\rm s} \sqrt{1-b^2}}{v} \ ,
\end{equation}

\noindent
where the orbital velocity is given as

\begin{equation}\label{eq:v}
v = \frac{2 \pi a}{P}
\end{equation}

\noindent
and $P$ is the orbital period. Together with an approximation for Kepler's third law, in which the cometary mass is vanishingly small compared to the stellar mass,

\begin{equation}\label{eq:P}
P \sim 2 \pi \sqrt{  \frac{a^3}{G M_{\rm s}} } \ ,
\end{equation}

\noindent
and where $G$ is the gravitational constant, we find

\begin{equation}\label{eq:dt}
{\Delta}t = 2 R_{\rm s} \sqrt{ \frac{a (1-b^2)}{G M_{\rm s}} }
\end{equation}

\noindent
or, equivalently,

\begin{equation}\label{eq:a}
a = \frac{G M_{\rm s}}{(1-b^2)} \left( \frac{{\Delta}t}{2 R_{\rm s}} \right) ^2 \ ,
\end{equation}

\noindent
where the transit impact parameter is

\begin{equation}\label{eq:b}
b = \frac{a}{R_{\rm s}} \tan{\Big (} \frac{\pi}{2} - i {\Big )} \ .
\end{equation}

\begin{figure}
    \centering
    \includegraphics[width=1\linewidth]{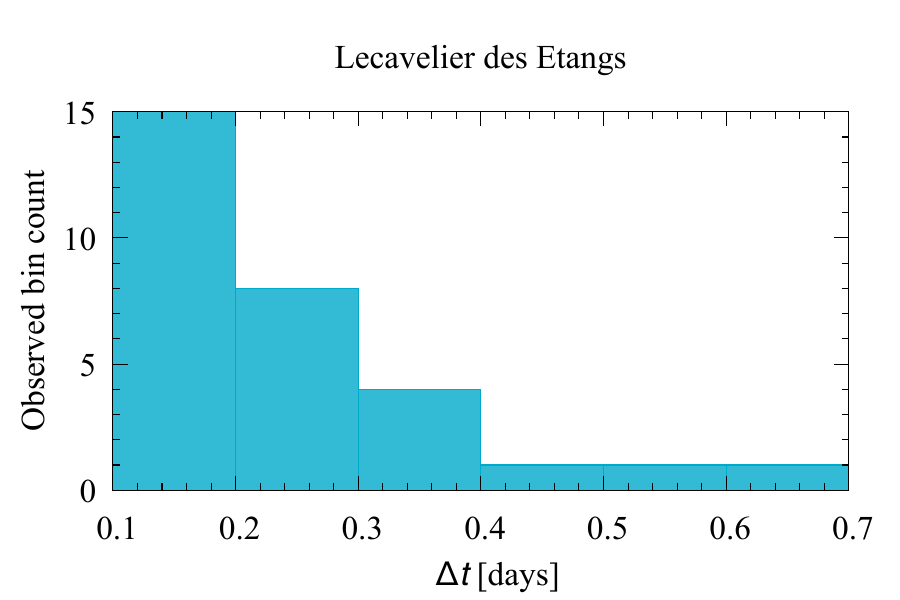}
    \caption{Transit durations of 30 exocometary events around $\beta$\,Pic as observed by \LecDE \, using TESS data.}
    \label{fig:LdE}
\end{figure}

\noindent
The geometrical transit probability ($\mathcal{P}$) favors the transits of objects in close orbits. Generally speaking, for an object that is much smaller than its host star $\mathcal{P} = R_{\rm s}/a$. And for reference, a planet at 1\,AU around a Sun-like star has a geometric transit probability of about 0.5\,\%. In our simulations, this selection effect is taken into account by randomizing the orbital inclination of the exocomets at their respective orbital distances to the star and by counting only those objects with transit impact parameters $b\,{\leq}\,1$. Strictly speaking it is really the probability density of $\cos(i)$ that is uniform. But our randomization of $0^\circ\,\leq\,i\,\leq\,90^\circ$ is still justified since transits only occur for inclinations near $90^\circ$, where $i$ and $\cos(i)$ converge.

In randomizing the orbital inclinations in Eq.~\eqref{eq:b} and assuming a distribution of exocometary orbits around $\beta$\,Pic we can {then} predict an observed distribution for ${\Delta}t$ and compare this prediction with the data of \LecDE \, compiled in Fig.~\ref{fig:LdE}.







\subsection{Hyperbolic orbits}

{While previous studies concluded that the observed transiting comets around $\beta$\,Pic have periastron distances near 0.1\,AU, comets are not
expected to be long-lived at such close-in separations from this A5 star for extended periods. Hence, the assumption of circular orbits is likely an oversimplification of the situation. Moreover, spectral absorption features of the cometary coma have shown in-transit acceleration, strongly suggesting significant orbital eccentricities \citep{2018MNRAS.479.1997K}. This is also in agreement with the orbital eccentricities of the solar system comets, which often have $e{\sim}1$.

We thus also explore the ${\Delta}t$ histograms for comets on hyperbolic orbits. Our nomenclature is illustrated together with a reference trajectory in Fig.~\ref{fig:hyperbolic}, where $\vv{r}$ is the radius vector of the comet, $\vv{v}$ the orbital velocity vector, $\phi$ the angle between $\vv{v}$ and the line perpendicular to $\vv{r}$, and $\theta$ the angle between the orientation of the periastron and our line of sight. What is relevant to our transit studies is the velocity component perpendicular to the line of sight, which we refer to as in-transit velocity ($v_{\rm t}$).

Our investigation of hyperbolic orbits has four free parameters, that is, $a$, $e$, $\theta$, and $i$. For any choice of $a$, $e$, and $\theta$, we compute

\begin{equation}\label{eq:phi_hyper}
\phi = \arctan \left( \frac{e \sin(\theta)}{1+e\cos(\theta)} \right) \ ,
\end{equation}

\begin{equation}\label{eq:r_hyper}
r = \frac{a(e^2-1)}{1+e\cos(\theta)} \ ,
\end{equation}

\begin{equation}\label{eq:v_hyper}
v = \sqrt{GM_{\rm s} \left( \frac{2}{r} - \frac{1}{a} \right)} \ ,
\end{equation}

\noindent
to finally obtain the in-transit velocity as $v_{\rm t} = v \cos(\phi)$, which we assume to be constant. The transit duration then follows as per Eq.~\eqref{eq:Dt}, but using $v_{\rm t}$ instead of $v$, and with $b$ as computed with Eq.~\eqref{eq:b} for any given choice of $a$ and $i$. Note that for hyperbolic orbits we use the convention of $a < 0$ to make Eqs.~\eqref{eq:phi_hyper}--\eqref{eq:v_hyper} applicable.

}

\begin{figure}
    \centering
    \includegraphics[width=0.58\linewidth]{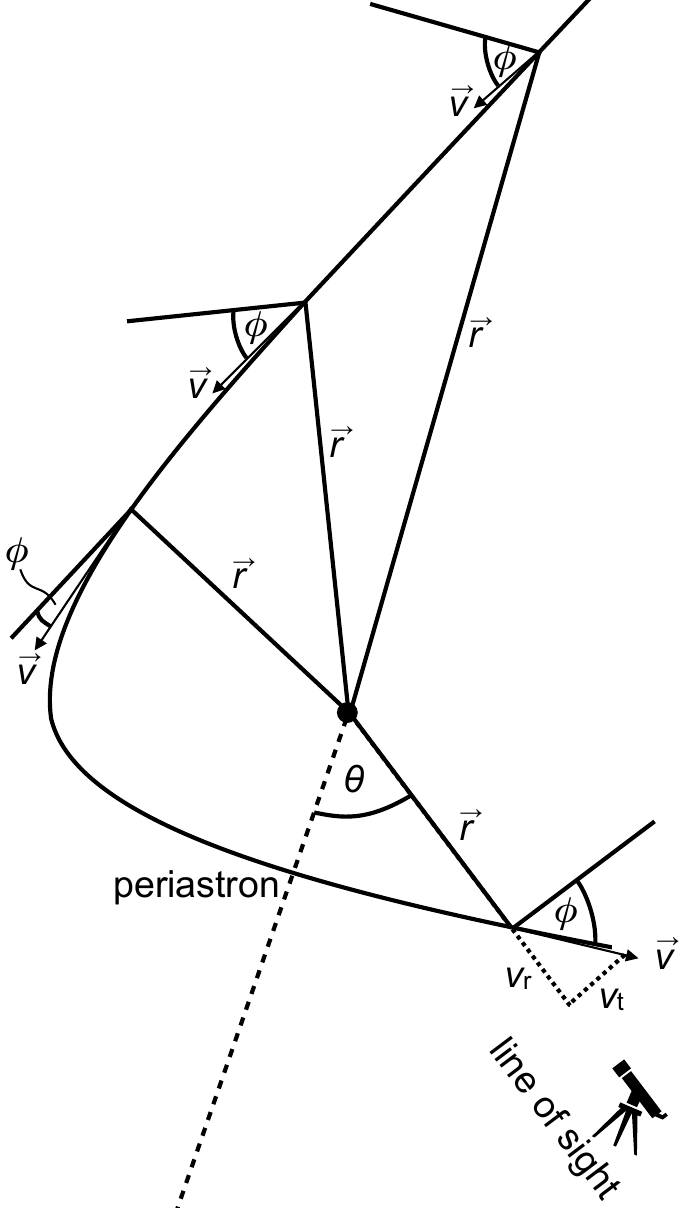}
    \caption{Parameterization of hyperbolic orbits.}
    \label{fig:hyperbolic}
\end{figure}

\begin{figure*}
    \centering
    \includegraphics[width=0.48\linewidth]{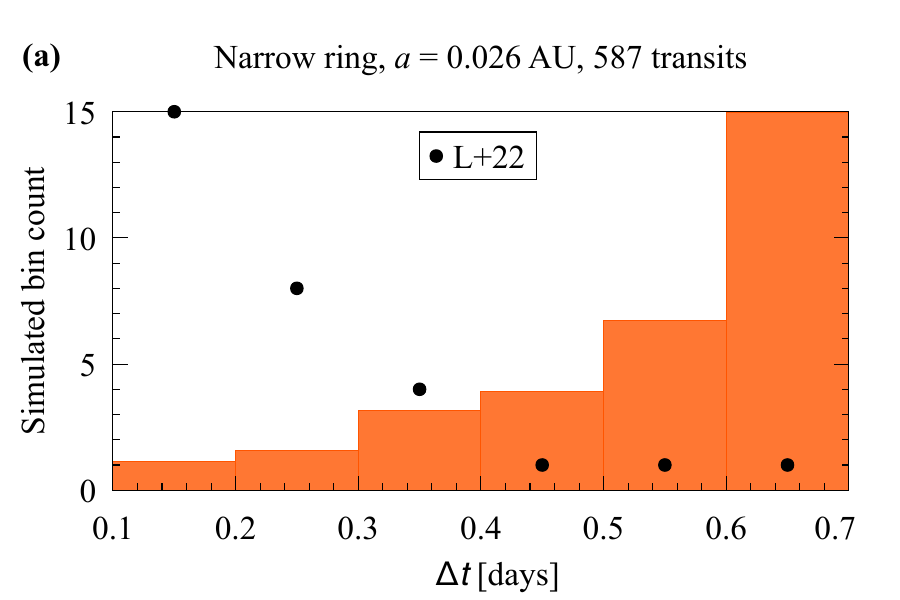}
    \hspace{0.25cm}
    \includegraphics[width=0.48\linewidth]{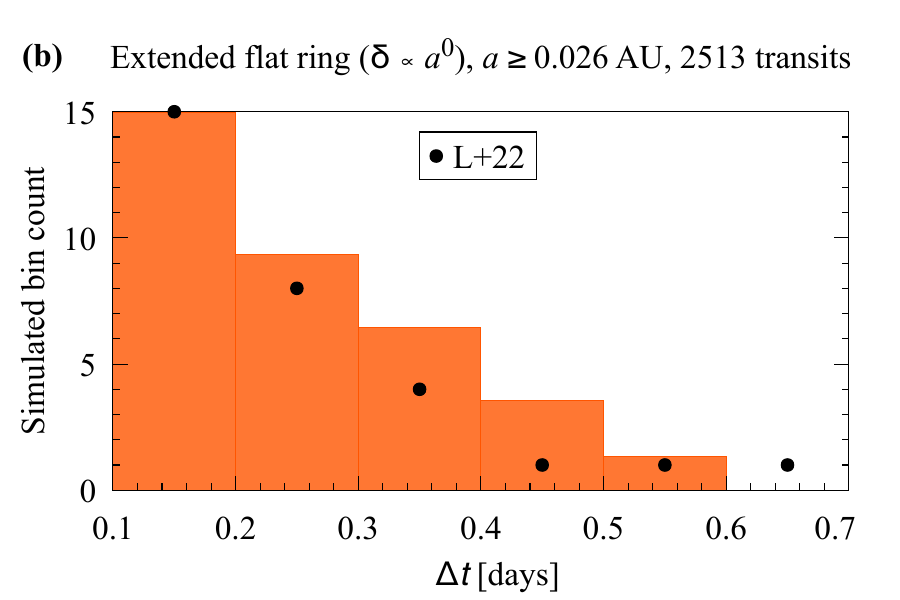}
    \caption{Histogram of the transit durations (Eq.~\ref{eq:dt}) based on 300,000 randomly chosen orbital orientations of an exocometary sample. Black dots illustrate the bin counts compiled from {\LecDE}. (a) The orbital distance was fixed to 2.5\,AU around $\beta$\,Pic. (b) The orbital distance was drawn from a flat distribution with $0.026\,{\rm AU}~{\leq}~a~{\leq}~2.5\,{\rm AU}$ around $\beta$\,Pic. Histograms have been normalized to 15 corresponding to Fig.~\ref{fig:LdE}.}
    \label{fig:a_fix_flat}
\end{figure*}

\section{Results}

For $\beta$\,Pic, we assume the best-fit radius of $R_{\rm s}=1.497\,R_\odot$ and a mass of $M_{\rm s}=1.797\,M_\odot$ (subscript $\odot$ referring to solar values) according to some of the most recent and well-constrained estimates using asteroseismology \citep{2019A&A...627A..28Z}. For the most short-duration transit events with ${\Delta}t=0.1$\,d and $b=0$, Eq.~\eqref{eq:a} for circular orbits then yields $a=0.026$\,AU. This is in agreement with the {estimated periastron distance} of 0.03\,AU by \LecDE \, although the authors used slightly different stellar mass and radius and a fixed mean transit path of $\bar{p}=\pi R_{\rm s}/2$. Taking $\bar{p}~=~2R_{\rm s}\sqrt{1-\bar{b}^2}$ in Eq.~\eqref{eq:p}, this is equivalent to a mean transit impact parameter of $\bar{b}~{\approx}~0.619$.

\begin{figure*}
    \centering
    \includegraphics[width=0.48\linewidth]{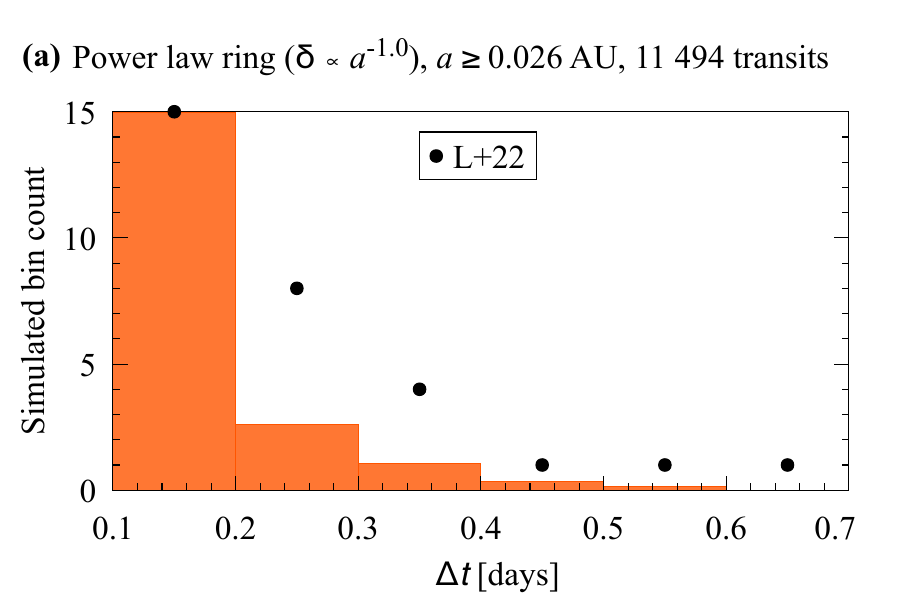}
    \hspace{0.25cm}
    \includegraphics[width=0.48\linewidth]{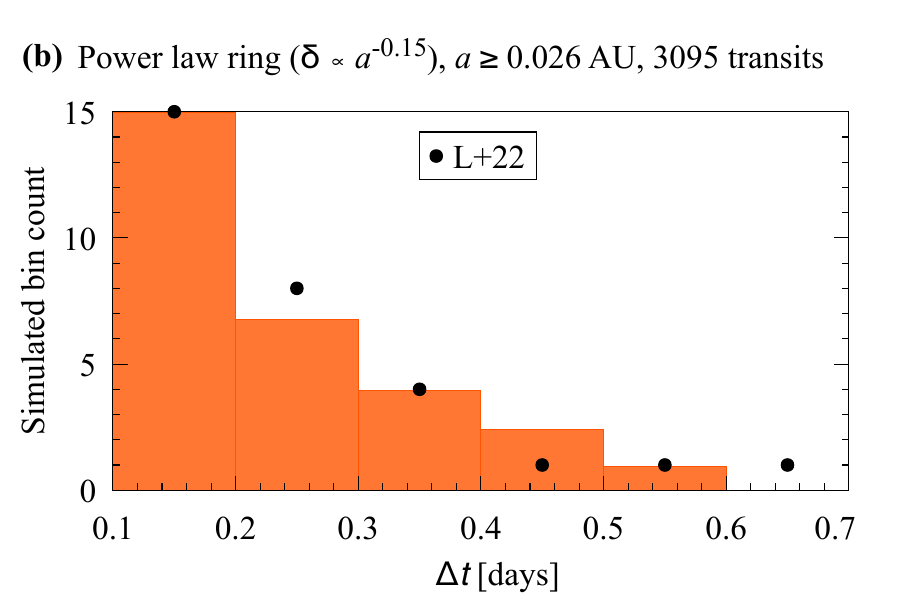}
    \caption{Similar to Fig.~\ref{fig:a_fix_flat} except that the orbital distances of the comets were drawn from different power law distributions. (a) The $a^{-1}$ power law with an inner truncation radius at $0.026\,{\rm AU}$ around $\beta$\,Pic. (a) The best-fit $a^{-0.15}$ power law with an inner truncation radius at $0.026\,{\rm AU}$ around $\beta$\,Pic.}
    \label{fig:a_power}
\end{figure*}

{Comets in such close-in orbits around $\beta$\,Pic, however, could hardly withstand evaporation for more than a few orbits. And so circular orbits are certainly an oversimplification of the real situation. Nevertheless, circular orbits, due to their geometric simplicity, are quite helpful to get a better understand of how different orbital distributions of the comets translate into different ${\Delta}t$ diagrams. Hence, we first explore several scenarios with circular orbits for didactic purpose, although the actual distribution of cometary trajectories around $\beta$\,Pic certainly looks different.}

\subsection{Narrow ring}

{First}, we execute simulations for a sample of exocomets at a fixed orbital distance. In this setup, the comet {surface} density ($\delta$) as a function of $a$ is zero everywhere except {for} an arbitrary distance $a'$. The exact value of $a'$ is irrelevant at this point because the shape of the resulting ${\Delta}t$ distribution depends only on the distribution of the transit impact parameters (Fig.~\ref{fig:b}) and normalization depends on the unknown occurrence rate of exocomet transits around $\beta$\,Pic. Moreover, as we will see, the shape of the histogram does not fit to the observations in the first place.

We thus take an arbitrary value of $a' = 2.5$\,AU and produce 300,000 random draws\footnote{The number of 300,000 realizations was chosen because this number is an integer multiple of orders of magnitudes higher (four in this case) than the 30 observations by \LecDE \, and computations on a standard laptop were executed in about 1\,s.} of the orbital inclination in the interval $0^\circ \leq i \leq 90^\circ$, the result of which is shown in Fig.~\ref{fig:a_fix_flat}(a). For better comparability to Fig.~\ref{fig:LdE} the histogram has been normalized to 15 along the ordinate, and the central values of the exocomet histogram from \LecDE \, have been added as black points. The total number of transits among the 300,000 realizations is 587, or 0.20\,\%.

Most important, and most obviously, the distribution in Fig.~\ref{fig:a_fix_flat}(a) shows a pile up towards high values of ${\Delta}t$ and a lack of short transit durations. This is in opposite to the histogram for the exocomets around $\beta$\,Pic (Fig.~\ref{fig:LdE}). Since the test objects in this simulation all have the same orbital semimajor axis, the shape of the histogram is entirely determined by the distribution of the transit impact parameters as described in Fig.~\ref{fig:b}. This discrepancy is not due an observational bias caused by the dependence of the geometric transit probability on the orbital semimajor axis. As a result, the stark discrepancy of the resulting distribution in Fig.~\ref{fig:a_fix_flat}(a) compared to the observations of \LecDE \, (Fig.~\ref{fig:LdE}) clearly shows that the ${\Delta}t$ distribution observed around $\beta$\,Pic cannot be explained by a narrow ring of comets at a given orbital distance with equally distributed transit impact parameters.

\subsection{Extended flat ring}

We move on and investigate an expanded ring of exocomets with a flat density distribution ($\delta \propto a^0 = {\rm const.}$) between 0.026\,AU, corresponding to the innermost orbit compatible with the observed minimum transit duration around $\beta$\,Pic, and 2.5\,AU. Initial tests showed that orbits beyond 2.5\,AU do not produce significant numbers of transiting objects, which is why this orbit was chosen as an outer boundary. The reason for this low rate of transits in wide orbits is due to the geometric transit probability, which scales as $\propto a^{-1}$.

The resulting ${\Delta}t$ histogram is shown in Fig.~\ref{fig:a_fix_flat}(b). The total number of transits among the 300,000 realizations is 2513, a fraction of 0.84\,\%. The key distinction with respect to the narrow ring scenario in Fig.~\ref{fig:a_fix_flat}(a) is the pile up towards short transit durations, which is in much better agreement with the observations. That said, with the histogram normalized to 15 transits in the [0.1\,d, 0.2\,d] bin, the extended flat ring scenario predicts a significant overabundance of transits for $0.2\,{\rm d} < {\Delta}t < 0.6\,{\rm d}$. This suggests that {in order to reproduce the observed ${\Delta}t$ distribution with circular orbits}, the comet density {of a hypothetical ring} around $\beta$\,Pic {would need to decline} with distance, which is what we explore in the following.

\begin{figure}
    \centering
    \includegraphics[width=1.0\linewidth]{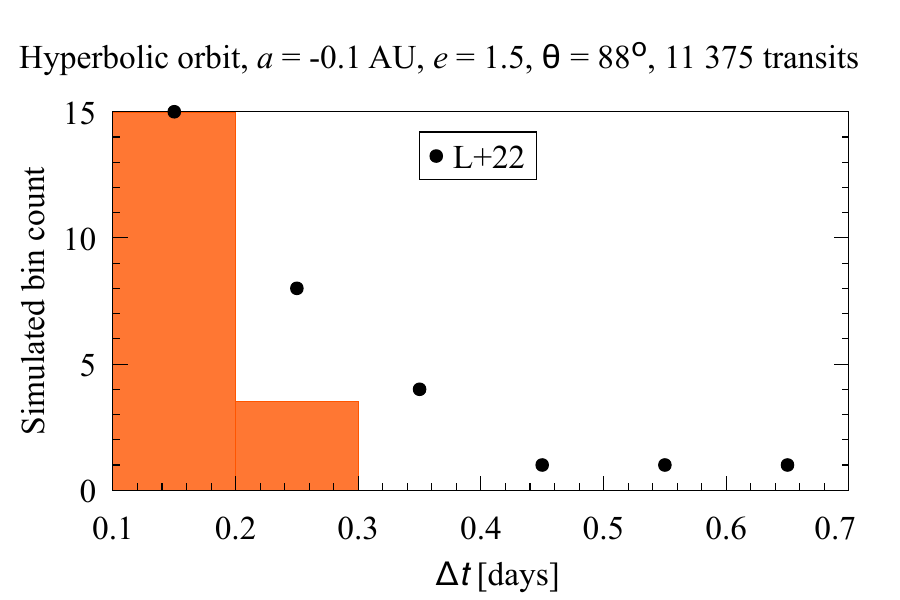}
    \caption{Histogram of the transit durations (Eq.~\ref{eq:dt}) based on 300,000 randomly chosen orbital orientations of an exocometary sample on a hyperbolic orbit. Black dots illustrate the bin counts compiled from {\LecDE}. The orbital elements are given in the title. The histogram has been normalized to 15 corresponding to Fig.~\ref{fig:LdE}.}
    \label{fig:hist_hyper}
\end{figure}

\subsection{Power law {ring}}

Next, we investigate an {inner ring boundary of 0.026\,AU (to reproduce the observed minimum transit durations with circular orbits) with an} exocomet density decline as per the power law $\delta \sim a^{-\beta}$. As an initial test, we look at a $\delta \propto a^{-1}$ power law ring, the resulting histogram of which is presented in Fig.~\ref{fig:a_power}(a). The total number of transits among the 300,000 realizations is 11\,494, corresponding to 3.83\,\%. When normalized to 15 transits in the [0.1\,d, 0.2\,d] bin, the theoretical distribution underestimates the observed transit bin counts around $\beta$\,Pic consistently for all of the longer transit durations. This result immediately shows that the exocomet density decline resulting from the initial guess of $\beta = -1$ is too steep.

To optimize our search for the best fitting value of $\beta$, we test a total of 40 power laws with $-2 \leq \beta \leq -0.05$ in steps of $0.05$. For each test value of $\beta$, we perform 300,000 randomized realizations, in which the orbital semimajor axis is chosen from the respective power law density and the transit geometry is drawn from a uniform distribution for $0^\circ \leq i \leq 90^\circ$. This results in a total of 1.2 million computer-generated realizations. For each of the 40 power laws, we calculate the residual sum of squares, which we find to be minimal for $\beta = -0.15$. Moreover, we compute the standard deviation of the best-fitting $\beta$ value from the $\chi^2$ distribution \citep{1986ApJ...305..740Z,2009A&A...496..191H} to obtain formal error bars, resulting in $\beta = -0.15_{-0.10}^{+0.05}$.

The resulting best-fit ${\Delta}t$ histogram for this $\delta \propto a^{-0.15}$ power law ring is presented in Fig.~\ref{fig:a_power}(b). It includes {3064} transits generated from 300,000 realization, corresponding to a transit fraction of {0.84}\,\%.

\subsection{Hyperbolic orbits}

Comets on circular orbits in extended flat rings or power law rings are very useful to get a sense of the parameter dependences and their effect on the ${\Delta}t$ histograms. But highly eccentric orbits are what we are ultimately interested in to find a physically plausible scenario. Comets will live longer on high eccentricity orbits, since most of the time is spent far from the star, and hence eccentric orbits are more physically plausible for a system of a given age.

We tested different hyperbolic orbits with various combinations of $a$, $e$, and $\theta$ with a flat prior on $i$ and still assuming negligible cometary radii with respect to the stellar diameter. Indeed, we did not find as good a match of the resulting ${\Delta}t$ distribution as we found for the power law ring. One of the best matches is shown in Fig.~\ref{fig:hist_hyper}, where $a = -0.1$\,AU, $e = 1.5$, and $\theta = 88^\circ$. In this setup, the cometary material is sweeping across the stellar disk as seen from Earth, when it is at a stellar distance of 0.12\,AU. In comparison to the ${\Delta}t$ histograms for circular orbits, however, we could not reproduce as wide a tail in the ${\Delta}t$ distribution. Parabolic orbits have the same problem of a relatively narrow ${\Delta}t$ distribution as shown in Fig.~\ref{fig:hist_hyper}.

In Fig.~\ref{fig:r_hyp} we illustrate the length of the star-comet radius vector ($r$) as a function of $\theta$. Solid lines represent near-parabolic orbits with $e=1.01$ for three choices of $a~\in~\{0.1, 0.2, 1.0\}$\,AU. For these near-parabolic orbits, most orientations of the periastron with respect to our line of sight do indeed result in relatively close-in orbital radii during the observed transits.

In hyperbolic orbits, $r$ is larger than in near-parabolic orbits for most values of $\theta$ (dashed lines for $e=1.2$), where cometary material could withstand evaporation. That being said, hyperbolic orbits have the issue that these objects would be external to the $\beta$\,Pic system. Moreover, none of our hyperbolic trajectory simulations with randomized orbital inclinations yields as good a match to the observations as the $\delta \propto a^{-0.15}$ power law ring. That being said, our tests were not exhaustive and a much better fit might exist. More work is also needed to explore the parameter space for a distribution of semi-major axes with parabolic orbits, since that would presumably help in much the same way as for circular orbits.

\begin{figure}
    \centering
    \includegraphics[width=1.\linewidth]{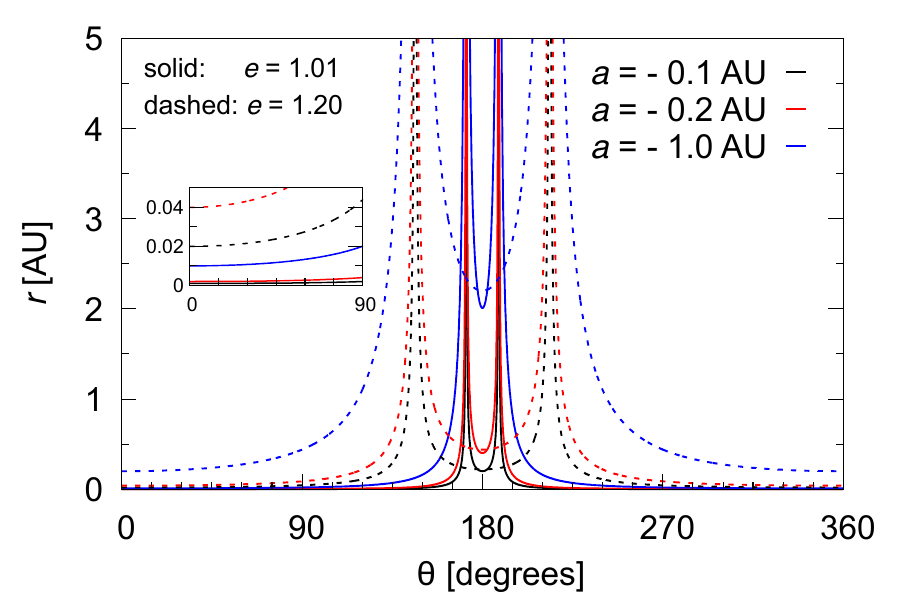}
    \caption{Variation of the orbital radius as a function of the orientation of the line of sight for hyperbolic orbits. Two different orbital eccentricities ($e \in \{1.01, 1.2 \}$)}
    \label{fig:r_hyp}
\end{figure}

\subsection{Effect of cometary size}
\label{sec:cometarysize}

Our calculations show that the observed transit durations of about 0.1\,d to 0.5\,d can be adequately explained by cometary material with periastron distances between 0.026\,AU and 0.12\,AU. In all these calculations we assume that the size of the comets is negligible for the transit duration.

If the observed transit durations were dominated by the extent of the transiting cometary material, then the time spent in transit would be much longer for any given instantaneous distance to the star. For a cloud of cometary material that has roughly the size of the star, for example, the transit duration would about double compared to our estimates with Eqs.~\eqref{eq:Dt} and \eqref{eq:dt}.

In turn, in order to fit the theoretical transit duration of an extended cometary cloud with a radius similar to $\beta$\,Pic to the observed transits durations of about 0.1\,d, the distance between the star and the comets would need to be about 0.006\,AU, which is physically implausible because this distance is smaller than the star.

Our results and interpretation are in agreement with simulations by \citet{1999A&A...343..916L} for a cometary transit around $\beta$\,Pic. These authors found that even a cloud of opaque cometary material around an evaporating cometary core is significantly smaller than the star. Their simulated transit light curve for a comet at a nominal distance of 1\,AU from the star \citep[][Fig.~1 therein]{1999A&A...343..916L} illustrates that most of the stellar occultation is completed after 15\,hr to 20\,hr, corresponding to about 0.6\, to 0.8\,d. With Eq.~\eqref{eq:dt} we obtain 0.6\,d.

\section{Conclusions}

Our analysis of the transit durations from a total of 30 exocometary transit events observed by \LecDE \, suggests that the density of cometary material ($\delta$) around $\beta$\,Pic can be reasonably well approximated by a power law distribution $\delta \sim a^{-\beta}$ with $\beta = -0.15_{-0.10}^{+0.05}$ under the assumption of circular orbits. This power law is derived from histograms of the exocometary transit durations that imply orbital separations of between 0.026 and about 2\,AU from star, details depending on the unknown transit impact parameters of the individual transit events. That being said, the short lifetimes that can be expected for comets closer than about 1\,AU from the star do not make this a very likely scenario. And so very recent origin of the observed comet family or families around $\beta$\,Pic would then need to be claimed.


A more plausible scenario involves comets on highly eccentric orbits, which we find could indeed lead to the observed transit durations of as short as about 0.1\,d. Our predicted ${\Delta}t$ histograms for a single family of comets on the same hyperbolic or near-parabolic orbit, however, cannot reproduce the observed transit times. Since we can also rule out that the cometary size is a dominant aspect, this means that either substantial fine tuning of the orientation of a highly eccentric orbit is required or that a recent breakup of an eccentric comet into a cometary family does a poor job of matching the data. Given that objects with similar orbits have long been used to explain that spectroscopic features around $\beta$\,Pic are normally redshifted \citet{1990A&A...236..202B} and given that a single comet family is a more natural model as it requires far fewer non-transiting objects than multiple families \citep{2016MNRAS.457.3988B, 2018MNRAS.473.5286W}, our conclusion might be quite powerful given the simplicity of the method applied in this paper.

Previous high-resolution stellar spectroscopy distinguished two families of comets located at $10\,({\pm}\,3)\,R_{\rm s}$ and $19\,({\pm}\,4)\,R_{\rm s}$ \citep{2014Natur.514..462K}, corresponding to about 0.07\,AU and 0.13\,AU respectively. The resulting transit durations would be about 0.16\,d and 0.22\,d and thus unresolved in our histograms. Nevertheless, our new results suggest that these previously detected cometary families are but the inner compact regions of what could really be an extended stream of comets on hyperbolic orbits. In other words, our results imply the existence of highly eccentric cometary material with a more extended range of semimajor axes than one expects just given the spectroscopic data.


Future photometric observations of exocometary transits around $\beta$\,Pic have the potential to further constrain the orbital distribution of comets around that star. In addition to the 30 transits analyzed in this report, there is the possibility that some very short (${\lesssim}\,0.1$\,d) events at the limit of detectability could still be found in the available data. It is well possible that TESS will re-observe $\beta$\,Pic in the future, which could add several dozen of additional transits. Key improvements on the data analysis side could be possible if the transit impact parameters of individual events can be measured. But this is challenging since it depends on proper modeling of the stellar limb darkening, and $\beta$\,Pic might be too photometrically active for this type is fine structure analysis.

{A natural extension of this work would be a systematic search for a best fitting solution of hyperbolic, near-parabolic orbits that reproduce the ${\Delta}t$ diagram inferred from \LecDE. Such an analysis might need to use a radially extended family of comets as suggested by our best-fitting $\delta \propto a^{-0.15}$ power law ring solution. And finally, our assumption of the comets being small compared to the star might be lifted and be replaced by a size-dependent comet distribution.}


\begin{acknowledgements}
The author is thankful to Alain Lecavelier des {\'E}tangs for comments on a draft of this manuscript. RH acknowledges support from the German Aerospace Agency (Deutsches Zentrum f\"ur Luft- und Raumfahrt) under PLATO Data Center grant 50OO1501.
\end{acknowledgements}

\bibliographystyle{aa}
\bibliography{literature}

\begin{thebibliography}{26}
\expandafter\ifx\csname natexlab\endcsname\relax\def\natexlab#1{#1}\fi

\bibitem[{{Beust} {et~al.}(1990){Beust}, {Lagrange-Henri}, {Vidal-Madjar}, \&
  {Ferlet}}]{1990A&A...236..202B}
{Beust}, H., {Lagrange-Henri}, A.~M., {Vidal-Madjar}, A., \& {Ferlet}, R. 1990,
  \aap, 236, 202

\bibitem[{{Boyajian} {et~al.}(2016){Boyajian}, {LaCourse}, {Rappaport},
  {Fabrycky}, {Fischer}, {Gandolfi}, {Kennedy}, {Korhonen}, {Liu}, {Moor},
  {Olah}, {Vida}, {Wyatt}, {Best}, {Brewer}, {Ciesla}, {Cs{\'a}k}, {Deeg},
  {Dupuy}, {Handler}, {Heng}, {Howell}, {Ishikawa}, {Kov{\'a}cs}, {Kozakis},
  {Kriskovics}, {Lehtinen}, {Lintott}, {Lynn}, {Nespral}, {Nikbakhsh},
  {Schawinski}, {Schmitt}, {Smith}, {Szabo}, {Szabo}, {Viuho}, {Wang},
  {Weiksnar}, {Bosch}, {Connors}, {Goodman}, {Green}, {Hoekstra}, {Jebson},
  {Jek}, {Omohundro}, {Schwengeler}, \& {Szewczyk}}]{2016MNRAS.457.3988B}
{Boyajian}, T.~S., {LaCourse}, D.~M., {Rappaport}, S.~A., {et~al.} 2016,
  \mnras, 457, 3988

\bibitem[{{Dohnanyi}(1969)}]{1969JGR....74.2531D}
{Dohnanyi}, J.~S. 1969, \jgr, 74, 2531

\bibitem[{{Ferlet} {et~al.}(1987){Ferlet}, {Hobbs}, \&
  {Vidal-Madjar}}]{1987A&A...185..267F}
{Ferlet}, R., {Hobbs}, L.~M., \& {Vidal-Madjar}, A. 1987, \aap, 185, 267

\bibitem[{{Gaia Collaboration} {et~al.}(2021){Gaia Collaboration}, {Brown},
  {Vallenari}, {Prusti}, {de Bruijne}, {Babusiaux}, {Biermann}, {Creevey},
  {Evans}, {Eyer}, {Hutton}, {Jansen}, {Jordi}, {Klioner}, {Lammers},
  {Lindegren}, {Luri}, {Mignard}, {Panem}, {Pourbaix}, {Randich}, {Sartoretti},
  {Soubiran}, {Walton}, {Arenou}, {Bailer-Jones}, {Bastian}, {Cropper},
  {Drimmel}, {Katz}, {Lattanzi}, {van Leeuwen}, {Bakker}, {Cacciari},
  {Casta{\~n}eda}, {De Angeli}, {Ducourant}, {Fabricius}, {Fouesneau},
  {Fr{\'e}mat}, {Guerra}, {Guerrier}, {Guiraud}, {Jean-Antoine Piccolo},
  {Masana}, {Messineo}, {Mowlavi}, {Nicolas}, {Nienartowicz}, {Pailler},
  {Panuzzo}, {Riclet}, {Roux}, {Seabroke}, {Sordo}, {Tanga}, {Th{\'e}venin},
  {Gracia-Abril}, {Portell}, {Teyssier}, {Altmann}, {Andrae}, {Bellas-Velidis},
  {Benson}, {Berthier}, {Blomme}, {Brugaletta}, {Burgess}, {Busso}, {Carry},
  {Cellino}, {Cheek}, {Clementini}, {Damerdji}, {Davidson}, {Delchambre},
  {Dell'Oro}, {Fern{\'a}ndez-Hern{\'a}ndez}, {Galluccio}, {Garc{\'\i}a-Lario},
  {Garcia-Reinaldos}, {Gonz{\'a}lez-N{\'u}{\~n}ez}, {Gosset}, {Haigron},
  {Halbwachs}, {Hambly}, {Harrison}, {Hatzidimitriou}, {Heiter},
  {Hern{\'a}ndez}, {Hestroffer}, {Hodgkin}, {Holl}, {Jan{\ss}en}, {Jevardat de
  Fombelle}, {Jordan}, {Krone-Martins}, {Lanzafame}, {L{\"o}ffler}, {Lorca},
  {Manteiga}, {Marchal}, {Marrese}, {Moitinho}, {Mora}, {Muinonen}, {Osborne},
  {Pancino}, {Pauwels}, {Petit}, {Recio-Blanco}, {Richards}, {Riello},
  {Rimoldini}, {Robin}, {Roegiers}, {Rybizki}, {Sarro}, {Siopis}, {Smith},
  {Sozzetti}, {Ulla}, {Utrilla}, {van Leeuwen}, {van Reeven}, {Abbas}, {Abreu
  Aramburu}, {Accart}, {Aerts}, {Aguado}, {Ajaj}, {Altavilla}, {{\'A}lvarez},
  {{\'A}lvarez Cid-Fuentes}, {Alves}, {Anderson}, {Anglada Varela}, {Antoja},
  {Audard}, {Baines}, {Baker}, {Balaguer-N{\'u}{\~n}ez}, {Balbinot}, {Balog},
  {Barache}, {Barbato}, {Barros}, {Barstow}, {Bartolom{\'e}}, {Bassilana},
  {Bauchet}, {Baudesson-Stella}, {Becciani}, {Bellazzini}, {Bernet}, {Bertone},
  {Bianchi}, {Blanco-Cuaresma}, {Boch}, {Bombrun}, {Bossini}, {Bouquillon},
  {Bragaglia}, {Bramante}, {Breedt}, {Bressan}, {Brouillet}, {Bucciarelli},
  {Burlacu}, {Busonero}, {Butkevich}, {Buzzi}, {Caffau}, {Cancelliere},
  {C{\'a}novas}, {Cantat-Gaudin}, {Carballo}, {Carlucci}, {Carnerero},
  {Carrasco}, {Casamiquela}, {Castellani}, {Castro-Ginard}, {Castro Sampol},
  {Chaoul}, {Charlot}, {Chemin}, {Chiavassa}, {Cioni}, {Comoretto}, {Cooper},
  {Cornez}, {Cowell}, {Crifo}, {Crosta}, {Crowley}, {Dafonte}, {Dapergolas},
  {David}, {David}, {de Laverny}, {De Luise}, {De March}, {De Ridder}, {de
  Souza}, {de Teodoro}, {de Torres}, {del Peloso}, {del Pozo}, {Delbo},
  {Delgado}, {Delgado}, {Delisle}, {Di Matteo}, {Diakite}, {Diener},
  {Distefano}, {Dolding}, {Eappachen}, {Edvardsson}, {Enke}, {Esquej}, {Fabre},
  {Fabrizio}, {Faigler}, {Fedorets}, {Fernique}, {Fienga}, {Figueras},
  {Fouron}, {Fragkoudi}, {Fraile}, {Franke}, {Gai}, {Garabato},
  {Garcia-Gutierrez}, {Garc{\'\i}a-Torres}, {Garofalo}, {Gavras}, {Gerlach},
  {Geyer}, {Giacobbe}, {Gilmore}, {Girona}, {Giuffrida}, {Gomel}, {Gomez},
  {Gonzalez-Santamaria}, {Gonz{\'a}lez-Vidal}, {Granvik},
  {Guti{\'e}rrez-S{\'a}nchez}, {Guy}, {Hauser}, {Haywood}, {Helmi}, {Hidalgo},
  {Hilger}, {H{\l}adczuk}, {Hobbs}, {Holland}, {Huckle}, {Jasniewicz},
  {Jonker}, {Juaristi Campillo}, {Julbe}, {Karbevska}, {Kervella}, {Khanna},
  {Kochoska}, {Kontizas}, {Kordopatis}, {Korn}, {Kostrzewa-Rutkowska},
  {Kruszy{\'n}ska}, {Lambert}, {Lanza}, {Lasne}, {Le Campion}, {Le Fustec},
  {Lebreton}, {Lebzelter}, {Leccia}, {Leclerc}, {Lecoeur-Taibi}, {Liao},
  {Licata}, {Lindstr{\o}m}, {Lister}, {Livanou}, {Lobel}, {Madrero Pardo},
  {Managau}, {Mann}, {Marchant}, {Marconi}, {Marcos Santos}, {Marinoni},
  {Marocco}, {Marshall}, {Martin Polo}, {Mart{\'\i}n-Fleitas}, {Masip},
  {Massari}, {Mastrobuono-Battisti}, {Mazeh}, {McMillan}, {Messina},
  {Michalik}, {Millar}, {Mints}, {Molina}, {Molinaro}, {Moln{\'a}r},
  {Montegriffo}, {Mor}, {Morbidelli}, {Morel}, {Morris}, {Mulone}, {Munoz},
  {Muraveva}, {Murphy}, {Musella}, {Noval}, {Ord{\'e}novic}, {Orr{\`u}},
  {Osinde}, {Pagani}, {Pagano}, {Palaversa}, {Palicio}, {Panahi}, {Pawlak},
  {Pe{\~n}alosa Esteller}, {Penttil{\"a}}, {Piersimoni}, {Pineau}, {Plachy},
  {Plum}, {Poggio}, {Poretti}, {Poujoulet}, {Pr{\v{s}}a}, {Pulone}, {Racero},
  {Ragaini}, {Rainer}, {Raiteri}, {Rambaux}, {Ramos}, {Ramos-Lerate}, {Re
  Fiorentin}, {Regibo}, {Reyl{\'e}}, {Ripepi}, {Riva}, {Rixon}, {Robichon},
  {Robin}, {Roelens}, {Rohrbasser}, {Romero-G{\'o}mez}, {Rowell}, {Royer},
  {Rybicki}, {Sadowski}, {Sagrist{\`a} Sell{\'e}s}, {Sahlmann}, {Salgado},
  {Salguero}, {Samaras}, {Sanchez Gimenez}, {Sanna}, {Santove{\~n}a},
  {Sarasso}, {Schultheis}, {Sciacca}, {Segol}, {Segovia}, {S{\'e}gransan},
  {Semeux}, {Shahaf}, {Siddiqui}, {Siebert}, {Siltala}, {Slezak}, {Smart},
  {Solano}, {Solitro}, {Souami}, {Souchay}, {Spagna}, {Spoto}, {Steele},
  {Steidelm{\"u}ller}, {Stephenson}, {S{\"u}veges}, {Szabados}, {Szegedi-Elek},
  {Taris}, {Tauran}, {Taylor}, {Teixeira}, {Thuillot}, {Tonello}, {Torra},
  {Torra}, {Turon}, {Unger}, {Vaillant}, {van Dillen}, {Vanel}, {Vecchiato},
  {Viala}, {Vicente}, {Voutsinas}, {Weiler}, {Wevers}, {Wyrzykowski}, {Yoldas},
  {Yvard}, {Zhao}, {Zorec}, {Zucker}, {Zurbach}, \&
  {Zwitter}}]{2021A&A...649A...1G}
{Gaia Collaboration}, {Brown}, A.~G.~A., {Vallenari}, A., {et~al.} 2021, \aap,
  649, A1

\bibitem[{{Gaia Collaboration} {et~al.}(2016){Gaia Collaboration}, {Prusti},
  {de Bruijne}, {Brown}, {Vallenari}, {Babusiaux}, {Bailer-Jones}, {Bastian},
  {Biermann}, {Evans}, {Eyer}, {Jansen}, {Jordi}, {Klioner}, {Lammers},
  {Lindegren}, {Luri}, {Mignard}, {Milligan}, {Panem}, {Poinsignon},
  {Pourbaix}, {Randich}, {Sarri}, {Sartoretti}, {Siddiqui}, {Soubiran},
  {Valette}, {van Leeuwen}, {Walton}, {Aerts}, {Arenou}, {Cropper}, {Drimmel},
  {H{\o}g}, {Katz}, {Lattanzi}, {O'Mullane}, {Grebel}, {Holland}, {Huc},
  {Passot}, {Bramante}, {Cacciari}, {Casta{\~n}eda}, {Chaoul}, {Cheek}, {De
  Angeli}, {Fabricius}, {Guerra}, {Hern{\'a}ndez}, {Jean-Antoine-Piccolo},
  {Masana}, {Messineo}, {Mowlavi}, {Nienartowicz}, {Ord{\'o}{\~n}ez-Blanco},
  {Panuzzo}, {Portell}, {Richards}, {Riello}, {Seabroke}, {Tanga},
  {Th{\'e}venin}, {Torra}, {Els}, {Gracia-Abril}, {Comoretto},
  {Garcia-Reinaldos}, {Lock}, {Mercier}, {Altmann}, {Andrae}, {Astraatmadja},
  {Bellas-Velidis}, {Benson}, {Berthier}, {Blomme}, {Busso}, {Carry},
  {Cellino}, {Clementini}, {Cowell}, {Creevey}, {Cuypers}, {Davidson}, {De
  Ridder}, {de Torres}, {Delchambre}, {Dell'Oro}, {Ducourant}, {Fr{\'e}mat},
  {Garc{\'\i}a-Torres}, {Gosset}, {Halbwachs}, {Hambly}, {Harrison}, {Hauser},
  {Hestroffer}, {Hodgkin}, {Huckle}, {Hutton}, {Jasniewicz}, {Jordan},
  {Kontizas}, {Korn}, {Lanzafame}, {Manteiga}, {Moitinho}, {Muinonen},
  {Osinde}, {Pancino}, {Pauwels}, {Petit}, {Recio-Blanco}, {Robin}, {Sarro},
  {Siopis}, {Smith}, {Smith}, {Sozzetti}, {Thuillot}, {van Reeven}, {Viala},
  {Abbas}, {Abreu Aramburu}, {Accart}, {Aguado}, {Allan}, {Allasia},
  {Altavilla}, {{\'A}lvarez}, {Alves}, {Anderson}, {Andrei}, {Anglada Varela},
  {Antiche}, {Antoja}, {Ant{\'o}n}, {Arcay}, {Atzei}, {Ayache}, {Bach},
  {Baker}, {Balaguer-N{\'u}{\~n}ez}, {Barache}, {Barata}, {Barbier}, {Barblan},
  {Baroni}, {Barrado y Navascu{\'e}s}, {Barros}, {Barstow}, {Becciani},
  {Bellazzini}, {Bellei}, {Bello Garc{\'\i}a}, {Belokurov}, {Bendjoya},
  {Berihuete}, {Bianchi}, {Bienaym{\'e}}, {Billebaud}, {Blagorodnova},
  {Blanco-Cuaresma}, {Boch}, {Bombrun}, {Borrachero}, {Bouquillon}, {Bourda},
  {Bouy}, {Bragaglia}, {Breddels}, {Brouillet}, {Br{\"u}semeister},
  {Bucciarelli}, {Budnik}, {Burgess}, {Burgon}, {Burlacu}, {Busonero}, {Buzzi},
  {Caffau}, {Cambras}, {Campbell}, {Cancelliere}, {Cantat-Gaudin}, {Carlucci},
  {Carrasco}, {Castellani}, {Charlot}, {Charnas}, {Charvet}, {Chassat},
  {Chiavassa}, {Clotet}, {Cocozza}, {Collins}, {Collins}, {Costigan}, {Crifo},
  {Cross}, {Crosta}, {Crowley}, {Dafonte}, {Damerdji}, {Dapergolas}, {David},
  {David}, {De Cat}, {de Felice}, {de Laverny}, {De Luise}, {De March}, {de
  Martino}, {de Souza}, {Debosscher}, {del Pozo}, {Delbo}, {Delgado},
  {Delgado}, {di Marco}, {Di Matteo}, {Diakite}, {Distefano}, {Dolding}, {Dos
  Anjos}, {Drazinos}, {Dur{\'a}n}, {Dzigan}, {Ecale}, {Edvardsson}, {Enke},
  {Erdmann}, {Escolar}, {Espina}, {Evans}, {Eynard Bontemps}, {Fabre},
  {Fabrizio}, {Faigler}, {Falc{\~a}o}, {Farr{\`a}s Casas}, {Faye}, {Federici},
  {Fedorets}, {Fern{\'a}ndez-Hern{\'a}ndez}, {Fernique}, {Fienga}, {Figueras},
  {Filippi}, {Findeisen}, {Fonti}, {Fouesneau}, {Fraile}, {Fraser}, {Fuchs},
  {Furnell}, {Gai}, {Galleti}, {Galluccio}, {Garabato}, {Garc{\'\i}a-Sedano},
  {Gar{\'e}}, {Garofalo}, {Garralda}, {Gavras}, {Gerssen}, {Geyer}, {Gilmore},
  {Girona}, {Giuffrida}, {Gomes}, {Gonz{\'a}lez-Marcos},
  {Gonz{\'a}lez-N{\'u}{\~n}ez}, {Gonz{\'a}lez-Vidal}, {Granvik}, {Guerrier},
  {Guillout}, {Guiraud}, {G{\'u}rpide}, {Guti{\'e}rrez-S{\'a}nchez}, {Guy},
  {Haigron}, {Hatzidimitriou}, {Haywood}, {Heiter}, {Helmi}, {Hobbs},
  {Hofmann}, {Holl}, {Holland}, {Hunt}, {Hypki}, {Icardi}, {Irwin}, {Jevardat
  de Fombelle}, {Jofr{\'e}}, {Jonker}, {Jorissen}, {Julbe}, {Karampelas},
  {Kochoska}, {Kohley}, {Kolenberg}, {Kontizas}, {Koposov}, {Kordopatis},
  {Koubsky}, {Kowalczyk}, {Krone-Martins}, {Kudryashova}, {Kull}, {Bachchan},
  {Lacoste-Seris}, {Lanza}, {Lavigne}, {Le Poncin-Lafitte}, {Lebreton},
  {Lebzelter}, {Leccia}, {Leclerc}, {Lecoeur-Taibi}, {Lemaitre}, {Lenhardt},
  {Leroux}, {Liao}, {Licata}, {Lindstr{\o}m}, {Lister}, {Livanou}, {Lobel},
  {L{\"o}ffler}, {L{\'o}pez}, {Lopez-Lozano}, {Lorenz}, {Loureiro},
  {MacDonald}, {Magalh{\~a}es Fernandes}, {Managau}, {Mann}, {Mantelet},
  {Marchal}, {Marchant}, {Marconi}, {Marie}, {Marinoni}, {Marrese},
  {Marschalk{\'o}}, {Marshall}, {Mart{\'\i}n-Fleitas}, {Martino}, {Mary},
  {Matijevi{\v{c}}}, {Mazeh}, {McMillan}, {Messina}, {Mestre}, {Michalik},
  {Millar}, {Miranda}, {Molina}, {Molinaro}, {Molinaro}, {Moln{\'a}r},
  {Moniez}, {Montegriffo}, {Monteiro}, {Mor}, {Mora}, {Morbidelli}, {Morel},
  {Morgenthaler}, {Morley}, {Morris}, {Mulone}, {Muraveva}, {Musella},
  {Narbonne}, {Nelemans}, {Nicastro}, {Noval}, {Ord{\'e}novic},
  {Ordieres-Mer{\'e}}, {Osborne}, {Pagani}, {Pagano}, {Pailler}, {Palacin},
  {Palaversa}, {Parsons}, {Paulsen}, {Pecoraro}, {Pedrosa}, {Pentik{\"a}inen},
  {Pereira}, {Pichon}, {Piersimoni}, {Pineau}, {Plachy}, {Plum}, {Poujoulet},
  {Pr{\v{s}}a}, {Pulone}, {Ragaini}, {Rago}, {Rambaux}, {Ramos-Lerate},
  {Ranalli}, {Rauw}, {Read}, {Regibo}, {Renk}, {Reyl{\'e}}, {Ribeiro},
  {Rimoldini}, {Ripepi}, {Riva}, {Rixon}, {Roelens}, {Romero-G{\'o}mez},
  {Rowell}, {Royer}, {Rudolph}, {Ruiz-Dern}, {Sadowski}, {Sagrist{\`a}
  Sell{\'e}s}, {Sahlmann}, {Salgado}, {Salguero}, {Sarasso}, {Savietto},
  {Schnorhk}, {Schultheis}, {Sciacca}, {Segol}, {Segovia}, {Segransan},
  {Serpell}, {Shih}, {Smareglia}, {Smart}, {Smith}, {Solano}, {Solitro},
  {Sordo}, {Soria Nieto}, {Souchay}, {Spagna}, {Spoto}, {Stampa}, {Steele},
  {Steidelm{\"u}ller}, {Stephenson}, {Stoev}, {Suess}, {S{\"u}veges}, {Surdej},
  {Szabados}, {Szegedi-Elek}, {Tapiador}, {Taris}, {Tauran}, {Taylor},
  {Teixeira}, {Terrett}, {Tingley}, {Trager}, {Turon}, {Ulla}, {Utrilla},
  {Valentini}, {van Elteren}, {Van Hemelryck}, {van Leeuwen}, {Varadi},
  {Vecchiato}, {Veljanoski}, {Via}, {Vicente}, {Vogt}, {Voss}, {Votruba},
  {Voutsinas}, {Walmsley}, {Weiler}, {Weingrill}, {Werner}, {Wevers},
  {Whitehead}, {Wyrzykowski}, {Yoldas}, {{\v{Z}}erjal}, {Zucker}, {Zurbach},
  {Zwitter}, {Alecu}, {Allen}, {Allende Prieto}, {Amorim},
  {Anglada-Escud{\'e}}, {Arsenijevic}, {Azaz}, {Balm}, {Beck}, {Bernstein},
  {Bigot}, {Bijaoui}, {Blasco}, {Bonfigli}, {Bono}, {Boudreault}, {Bressan},
  {Brown}, {Brunet}, {Bunclark}, {Buonanno}, {Butkevich}, {Carret}, {Carrion},
  {Chemin}, {Ch{\'e}reau}, {Corcione}, {Darmigny}, {de Boer}, {de Teodoro}, {de
  Zeeuw}, {Delle Luche}, {Domingues}, {Dubath}, {Fodor}, {Fr{\'e}zouls},
  {Fries}, {Fustes}, {Fyfe}, {Gallardo}, {Gallegos}, {Gardiol}, {Gebran},
  {Gomboc}, {G{\'o}mez}, {Grux}, {Gueguen}, {Heyrovsky}, {Hoar}, {Iannicola},
  {Isasi Parache}, {Janotto}, {Joliet}, {Jonckheere}, {Keil}, {Kim},
  {Klagyivik}, {Klar}, {Knude}, {Kochukhov}, {Kolka}, {Kos}, {Kutka}, {Lainey},
  {LeBouquin}, {Liu}, {Loreggia}, {Makarov}, {Marseille}, {Martayan},
  {Martinez-Rubi}, {Massart}, {Meynadier}, {Mignot}, {Munari}, {Nguyen},
  {Nordlander}, {Ocvirk}, {O'Flaherty}, {Olias Sanz}, {Ortiz}, {Osorio},
  {Oszkiewicz}, {Ouzounis}, {Palmer}, {Park}, {Pasquato}, {Peltzer}, {Peralta},
  {P{\'e}turaud}, {Pieniluoma}, {Pigozzi}, {Poels}, {Prat}, {Prod'homme},
  {Raison}, {Rebordao}, {Risquez}, {Rocca-Volmerange}, {Rosen}, {Ruiz-Fuertes},
  {Russo}, {Sembay}, {Serraller Vizcaino}, {Short}, {Siebert}, {Silva},
  {Sinachopoulos}, {Slezak}, {Soffel}, {Sosnowska}, {Strai{\v{z}}ys}, {ter
  Linden}, {Terrell}, {Theil}, {Tiede}, {Troisi}, {Tsalmantza}, {Tur},
  {Vaccari}, {Vachier}, {Valles}, {Van Hamme}, {Veltz}, {Virtanen}, {Wallut},
  {Wichmann}, {Wilkinson}, {Ziaeepour}, \& {Zschocke}}]{2016A&A...595A...1G}
{Gaia Collaboration}, {Prusti}, T., {de Bruijne}, J.~H.~J., {et~al.} 2016,
  \aap, 595, A1

\bibitem[{{Heller} {et~al.}(2009){Heller}, {Homeier}, {Dreizler}, \&
  {{\O}stensen}}]{2009A&A...496..191H}
{Heller}, R., {Homeier}, D., {Dreizler}, S., \& {{\O}stensen}, R. 2009, \aap,
  496, 191

\bibitem[{{Kennedy}(2018)}]{2018MNRAS.479.1997K}
{Kennedy}, G.~M. 2018, \mnras, 479, 1997

\bibitem[{{Kenworthy} {et~al.}(2021){Kenworthy}, {Mellon}, {Bailey}, {Stuik},
  {Dorval}, {Talens}, {Crawford}, {Mamajek}, {Laginja}, {Ireland}, {Lomberg},
  {Kuhn}, {Snellen}, {Zwintz}, {Kuschnig}, {Kennedy}, {Abe}, {Agabi},
  {Mekarnia}, {Guillot}, {Schmider}, {Stee}, {de Pra}, {Buttu}, {Crouzet},
  {Kalas}, {Wang}, {Stevenson}, {de Mooij}, {Lagrange}, {Lacour}, {Lecavelier
  des Etangs}, {Nowak}, {Str{\o}m}, {Hui}, \& {Wang}}]{2021A&A...648A..15K}
{Kenworthy}, M.~A., {Mellon}, S.~N., {Bailey}, J.~I., {et~al.} 2021, \aap, 648,
  A15

\bibitem[{{Kiefer} {et~al.}(2014){Kiefer}, {Lecavelier des Etangs}, {Boissier},
  {Vidal-Madjar}, {Beust}, {Lagrange}, {H{\'e}brard}, \&
  {Ferlet}}]{2014Natur.514..462K}
{Kiefer}, F., {Lecavelier des Etangs}, A., {Boissier}, J., {et~al.} 2014, \nat,
  514, 462

\bibitem[{{Lagrange} {et~al.}(2010){Lagrange}, {Bonnefoy}, {Chauvin}, {Apai},
  {Ehrenreich}, {Boccaletti}, {Gratadour}, {Rouan}, {Mouillet}, {Lacour}, \&
  {Kasper}}]{2010Sci...329...57L}
{Lagrange}, A.~M., {Bonnefoy}, M., {Chauvin}, G., {et~al.} 2010, Science, 329,
  57

\bibitem[{{Lagrange} {et~al.}(2009){Lagrange}, {Gratadour}, {Chauvin}, {Fusco},
  {Ehrenreich}, {Mouillet}, {Rousset}, {Rouan}, {Allard}, {Gendron}, {Charton},
  {Mugnier}, {Rabou}, {Montri}, \& {Lacombe}}]{2009A&A...493L..21L}
{Lagrange}, A.~M., {Gratadour}, D., {Chauvin}, G., {et~al.} 2009, \aap, 493,
  L21

\bibitem[{{Lagrange} {et~al.}(2019){Lagrange}, {Meunier}, {Rubini}, {Keppler},
  {Galland}, {Chapellier}, {Michel}, {Balona}, {Beust}, {Guillot}, {Grandjean},
  {Borgniet}, {M{\'e}karnia}, {Wilson}, {Kiefer}, {Bonnefoy}, {Lillo-Box},
  {Pantoja}, {Jones}, {Iglesias}, {Rodet}, {Diaz}, {Zapata}, {Abe}, \&
  {Schmider}}]{2019NatAs...3.1135L}
{Lagrange}, A.~M., {Meunier}, N., {Rubini}, P., {et~al.} 2019, Nature
  Astronomy, 3, 1135

\bibitem[{{Lecavelier des Etangs} {et~al.}(2022){Lecavelier des Etangs},
  {Cros}, {H{\'e}brard}, {Martioli}, {Duquesnoy}, {Kenworthy}, {Kiefer},
  {Lacour}, {Lagrange}, {Meunier}, \& {Vidal-Madjar}}]{LdE2022}
{Lecavelier des Etangs}, A., {Cros}, L., {H{\'e}brard}, G., {et~al.} 2022,
  Scientific Reports, 12, 5855

\bibitem[{{Lecavelier Des Etangs} {et~al.}(1995){Lecavelier Des Etangs},
  {Deleuil}, {Vidal-Madjar}, {Ferlet}, {Nitschelm}, {Nicolet}, \&
  {Lagrange-Henri}}]{1995A&A...299..557L}
{Lecavelier Des Etangs}, A., {Deleuil}, M., {Vidal-Madjar}, A., {et~al.} 1995,
  \aap, 299, 557

\bibitem[{{Lecavelier Des Etangs} {et~al.}(1999){Lecavelier Des Etangs},
  {Vidal-Madjar}, \& {Ferlet}}]{1999A&A...343..916L}
{Lecavelier Des Etangs}, A., {Vidal-Madjar}, A., \& {Ferlet}, R. 1999, \aap,
  343, 916

\bibitem[{{Mamajek} \& {Bell}(2014)}]{2014MNRAS.445.2169M}
{Mamajek}, E.~E. \& {Bell}, C. P.~M. 2014, \mnras, 445, 2169

\bibitem[{{Miret-Roig} {et~al.}(2020){Miret-Roig}, {Galli}, {Brandner}, {Bouy},
  {Barrado}, {Olivares}, {Antoja}, {Romero-G{\'o}mez}, {Figueras}, \&
  {Lillo-Box}}]{2020A&A...642A.179M}
{Miret-Roig}, N., {Galli}, P.~A.~B., {Brandner}, W., {et~al.} 2020, \aap, 642,
  A179

\bibitem[{{Nowak} {et~al.}(2020){Nowak}, {Lacour}, {Lagrange}, {Rubini},
  {Wang}, {Stolker}, {Abuter}, {Amorim}, {Asensio-Torres}, {Baub{\"o}ck},
  {Benisty}, {Berger}, {Beust}, {Blunt}, {Boccaletti}, {Bonnefoy}, {Bonnet},
  {Brandner}, {Cantalloube}, {Charnay}, {Choquet}, {Christiaens}, {Cl{\'e}net},
  {Coud{\'e} Du Foresto}, {Cridland}, {de Zeeuw}, {Dembet}, {Dexter},
  {Drescher}, {Duvert}, {Eckart}, {Eisenhauer}, {Gao}, {Garcia}, {Garcia
  Lopez}, {Gardner}, {Gendron}, {Genzel}, {Gillessen}, {Girard}, {Grandjean},
  {Haubois}, {Hei{\ss}el}, {Henning}, {Hinkley}, {Hippler}, {Horrobin},
  {Houll{\'e}}, {Hubert}, {Jim{\'e}nez-Rosales}, {Jocou}, {Kammerer},
  {Kervella}, {Keppler}, {Kreidberg}, {Kulikauskas}, {Lapeyr{\`e}re}, {Le
  Bouquin}, {L{\'e}na}, {M{\'e}rand}, {Maire}, {Molli{\`e}re}, {Monnier},
  {Mouillet}, {M{\"u}ller}, {Nasedkin}, {Ott}, {Otten}, {Paumard}, {Paladini},
  {Perraut}, {Perrin}, {Pueyo}, {Pfuhl}, {Rameau}, {Rodet},
  {Rodr{\'\i}guez-Coira}, {Rousset}, {Scheithauer}, {Shangguan}, {Stadler},
  {Straub}, {Straubmeier}, {Sturm}, {Tacconi}, {van Dishoeck}, {Vigan},
  {Vincent}, {von Fellenberg}, {Ward-Duong}, {Widmann}, {Wieprecht},
  {Wiezorrek}, {Woillez}, \& {Gravity Collaboration}}]{2020A&A...642L...2N}
{Nowak}, M., {Lacour}, S., {Lagrange}, A.~M., {et~al.} 2020, \aap, 642, L2

\bibitem[{{Pavlenko} {et~al.}(2022){Pavlenko}, {Kulyk}, {Shubina}, {Vasylenko},
  {Dobrycheva}, \& {Korsun}}]{2022A&A...660A..49P}
{Pavlenko}, Y., {Kulyk}, I., {Shubina}, O., {et~al.} 2022, \aap, 660, A49

\bibitem[{{Rappaport} {et~al.}(2018){Rappaport}, {Vanderburg}, {Jacobs},
  {LaCourse}, {Jenkins}, {Kraus}, {Rizzuto}, {Latham}, {Bieryla}, {Lazarevic},
  \& {Schmitt}}]{2018MNRAS.474.1453R}
{Rappaport}, S., {Vanderburg}, A., {Jacobs}, T., {et~al.} 2018, \mnras, 474,
  1453

\bibitem[{{Smith} \& {Terrile}(1984)}]{1984Sci...226.1421S}
{Smith}, B.~A. \& {Terrile}, R.~J. 1984, Science, 226, 1421

\bibitem[{{Wyatt} {et~al.}(2018){Wyatt}, {van Lieshout}, {Kennedy}, \&
  {Boyajian}}]{2018MNRAS.473.5286W}
{Wyatt}, M.~C., {van Lieshout}, R., {Kennedy}, G.~M., \& {Boyajian}, T.~S.
  2018, \mnras, 473, 5286

\bibitem[{{Zhang} {et~al.}(1986){Zhang}, {Robinson}, \&
  {Nather}}]{1986ApJ...305..740Z}
{Zhang}, E.~H., {Robinson}, E.~L., \& {Nather}, R.~E. 1986, \apj, 305, 740

\bibitem[{{Zieba} {et~al.}(2019){Zieba}, {Zwintz}, {Kenworthy}, \&
  {Kennedy}}]{2019A&A...625L..13Z}
{Zieba}, S., {Zwintz}, K., {Kenworthy}, M.~A., \& {Kennedy}, G.~M. 2019, \aap,
  625, L13

\bibitem[{{Zwintz} {et~al.}(2019){Zwintz}, {Reese}, {Neiner}, {Pigulski},
  {Kuschnig}, {M{\"u}llner}, {Zieba}, {Abe}, {Guillot}, {Handler}, {Kenworthy},
  {Stuik}, {Moffat}, {Popowicz}, {Rucinski}, {Wade}, {Weiss}, {Bailey},
  {Crawford}, {Ireland}, {Lomberg}, {Mamajek}, {Mellon}, \&
  {Talens}}]{2019A&A...627A..28Z}
{Zwintz}, K., {Reese}, D.~R., {Neiner}, C., {et~al.} 2019, \aap, 627, A28

\end{thebibliography}

\end{document}